\documentstyle[aps,prl,multicol,epsf]{revtex}
\renewcommand{\v}[1]{{\bf #1}}
\renewcommand{\narrowtext}{\begin{multicols}{2} \global\columnwidth20.5pc}
\renewcommand{\widetext}{\end{multicols} \global\columnwidth42.5pc}
\renewcommand{\narrowtext}{\begin{multicols}{2} \global\columnwidth20.5pc}
\renewcommand{\widetext}{\end{multicols} \global\columnwidth42.5pc}

\begin{document}
\renewcommand{\v}[1]{{\bf #1}}
\newcommand{\be}{\begin{equation}}
\newcommand{\ee}{\end{equation}}
\newcommand{\ba}{\begin{eqnarray}}
\newcommand{\ea}{\end{eqnarray}}
\newcommand{\nn}{\nonumber \\}
\newcommand{\bbar}{b^{+}}
\newcommand{\cbar}{c^{+}}
\newcommand{\dbar}{d^{+}}
\newcommand{\fbar}{f^+}
\newcommand{\ed}{\epsilon_d}
\newcommand{\ek}{\epsilon_k}
\newcommand{\mul}{\mu_L}
\newcommand{\mur}{\mu_R}
\newcommand{\bl}{b_1}
\newcommand{\br}{b_2}
\newcommand{\bbarl}{b^+_1}
\newcommand{\bbarr}{b^+_2}
\newcommand{\dbarsigma}{d^+_\sigma}
\newcommand{\flsigma}{f_{1\sigma}}
\newcommand{\frsigma}{f_{2\sigma}}
\newcommand{\fsigma}{f_\sigma}
\newcommand{\fbarsigma}{f^+_\sigma}
\newcommand{\fbarlsigma}{f^+_{1\sigma}}
\newcommand{\fbarrsigma}{f^+_{2\sigma}}
\newcommand{\lambdal}{\lambda_1}
\newcommand{\lambdar}{\lambda_2}
\newcommand{\gammal}{\Gamma_L}
\newcommand{\gammar}{\Gamma_R}
\newcommand{\tstar}{T_K^*}
\title{ A New Mean-Field Theory of the Kondo Resonance at Finite Bias}

\author{Jung Hoon Han}
\address{Department of Physics, SungKyunKwan University,
Suwon 440-746, Korea}
\maketitle
\draft

\begin{abstract}
We introduce a new slave-boson mean-field treatment of the Kondo
effect in a quantum dot attached to the leads, when the bias
voltage across the leads is finite. The model employs two slave
boson and two pseudo-fermion operators to express the localized
electron. The solution of the mean-field equations gives in
general two resonance peaks pinned to the chemical potential of
each lead. The Kondo temperature is shown to scale as $min(\tstar,
(\tstar)^2/V)$, where $\tstar$ is the Kondo temperature at
equilibrium, and $V$ is the chemical potential difference of the
leads.
\end{abstract}

\narrowtext The realization of  Kondo resonance in a single
molecule coupled to metallic leads offers a prime example of the
many-body correlation effect at work on the nanometer
scale\cite{molecular-kondo}. The rapid advance in the fabrication
technology of molecular-scale systems with current-carrying
capability will produce growing opportunity to observe Kondo
effect in a more intricate setup. On the theoretical side, it is
important to develop tools suitable to describe Kondo phenomena in
situations of increasing complexity.

The Kondo effect in a quantum dot coupled to metallic reservoir
has been thoroughly studied, and many theoretical tools are
available\cite{meir-wingreen}. The formation of a resonance state
at low temperature (Abrikosov-Suhl resonance) at the Fermi level
of the host metal can be understood as a consequence of
hybridization between a localized electron and the host electrons.
The slave-boson theory captures many of the essential features of
the Kondo resonance in the so-called Fermi-liquid
regime\cite{hewson}. For a quantum dot coupled to leads with
different chemical potentials, {\it i.e.} $\mu_L \neq \mu_R$ where
$\mu_L$ and $\mu_R$ are chemical potentials of the left and the
right leads, however, there is a discrepancy in the outcome of the
slave-boson mean-field theory(SBMFT) (predicting a single Kondo
resonance)\cite{langreth} and that of equation-of-motion(EOM) or
non-crossing-approximation(NCA) theories (predicting a splitting
of Kondo resonances) \cite{meir-wingreen}.

It is important to resolve this discrepancy fully, in view of the
simplicity (hence its utility) of SBMFT with which to capture
essential features of the Kondo physics. It will be desirable if
the outcome of the mean-field theory can be made to agree with
that of the more advanced techniques such as NCA or EOM methods.
In this paper, we assume that the predictions of the NCA/EOM is
indeed correct\cite{experiment}, and offer a new version of SBMFT
that gives rise to split Kondo peaks. We conclude, on the basis of
the new mean-field theory, that (1) the spectral function of the
quantum dot has resonances pinned at the chemical potentials of
each lead, (2) the Kondo temperature scales as $min (\tstar,
(\tstar)^2/V)$ where $\tstar$ is the Kondo temperature in the case
of un-biased leads coupled to the dot, and $V$ is the chemical
potential difference of the leads. In the following we present the
mathematical formulation of the theory and discuss physical
consequences of the solution that emerges.

A quantum dot coupled to two leads via tunnelling may be modelled
by the Hamiltonian \be H = H_L + H_R + H_d + H_T \ee where the
lead, the dot, and the tunnelling parts are \ba H_{L(R)}
&=&\sum_{k\in L(R),\sigma} (\epsilon_k -\mu_{L(R)})\cbar_{L(R)
k\sigma}c_{L(R) k\sigma} \nn H_d &=& -\ed \sum_\sigma
\dbar_{\sigma}d_{\sigma} + U n_d (n_d -1) \nn H_T &= &V_L
\sum_{k\in L,\sigma} \cbar_{Lk\sigma} d_\sigma +V_R \sum_{k\in
R,\sigma} \cbar_{Rk\sigma} d_\sigma+
h.c.\label{kondo-H}\label{original-H}\ea Tunnelling amplitude to
each lead is given by $V_L$ and $V_R$ while the density of states
of leads are assumed to be constant, $\rho_L$ and $\rho_R$
respectively. We consider the case where the bare energy level of
the dot $-\ed (<0)$ and the energy of double occupancy $U-\ed
(>0)$ are both large (Kondo regime).

The $d$-level operator $\dbarsigma$ in the $U\rightarrow\infty$
limit is really $\dbarsigma=|\sigma\rangle\langle e |$, taking an
empty state ($|e\rangle$) into a singly occupied state of spin
$\sigma$ ( $|\sigma\rangle)$, because doubly-occupied state is
energetically forbidden. Commutation relations for $d_\sigma$ thus
differ from the canonical one: \be \{d_\sigma,
\dbarsigma\}=1-d^+_{-\sigma}d_{-\sigma}, ~~~~ \{d_\sigma,
d^+_{-\sigma}\}=d^+_{-\sigma}d_{\sigma}.\label{restricted-comm}\ee
In the usual slave-boson approach, the localized electron operator
$d_\sigma$ is replaced by $b^+ f_\sigma$ with $b$ and $f_\sigma$
being the standard boson and fermion operators. Commutation
relations, Eq. (\ref{restricted-comm}), are recovered provided we
require $\bbar b +\sum_\sigma \fbarsigma \fsigma =1$. The
$d_\sigma$ operators in the Hamiltonian are replaced by an
equivalent expression, $d_\sigma \rightarrow \bbar \fsigma$, while
the constraint is included in the form of a Lagrange multiplier,
\be H \rightarrow H'= H+\lambda (\bbar b +\sum_\sigma \fbar_\sigma
f_\sigma -1).\ee The free parameters $\lambda$ and $b$ are
self-consistently determined by the condition \be {\partial
\langle H' \rangle \over
\partial \lambda} = 0 = {\partial \langle H' \rangle \over
\partial b}.\label{stationary-state-eq}\ee
The stationary state solution describes a narrow resonance state
formed at the chemical potential of the lead and gives a
quantitative estimate of the width of the resonance that is also
associated with the Kondo temperature $\tstar$.

The same approach was adapted by Aguado and Langreth (AL) to the
non-equilibrium case, $\mu_L \neq\mu_R$\cite{langreth}. One finds
in this case that a single Kondo resonance state is formed at the
energy level between the chemical potentials of each lead. In
physical terms, the chemical potentials of the leads are
``averaged over", before an interaction with the localized
electron is considered.

The root of the Abrikosov-Suhl resonance is the high-order virtual
hopping process that in effect exchanges spins between the lead
electrons and the localized electron. When more than one lead is
attached to a dot, it appears that a correct picture is to view
such hopping processes as occurring independently with respect to
each lead, as far as one can ignore a direct interaction between
the leads. Naturally, then, the resonance state will have to be
formed for each lead, at the chemical potential of the
corresponding lead. Motivated by this picture, let us write the
$d$-level electron operator as \be d_\sigma = \alpha\bbarl
\flsigma + \beta\bbarr \frsigma.\label{dsigma}\ee Two sets of
slave-boson ($\bl, \br$) and pseudo-fermion ($\flsigma, \frsigma$)
operators have been introduced, with relative weights $\alpha$ and
$\beta$ to be determined shortly\cite{N-leads}. The commutation
algebra, Eq. (\ref{restricted-comm}), are recovered provided we
require the following: \ba (a) &&~~ \bbarl \bl + \sum_\sigma
\fbarlsigma \flsigma = p \nn (b) &&~~ \bbarr \br + \sum_\sigma
\fbarrsigma \frsigma = q \nn (c) &&~~ \dbarsigma d_\sigma =
\alpha^2 \fbarlsigma \flsigma + \beta^2 \fbarrsigma \frsigma \nn
(d) && ~~\alpha^2 p + \beta^2 q = 1.\label{condition1}\ea Two
positive constants $p$ and $q$ have been introduced in the above
relations. The empty state $|e\rangle$ takes on the
representation,

\be |e\rangle = (m \bbarl \bbarr +
{n\over\sqrt{2}}(\bbarl)^2)|0\rangle,\label{emptystate}\ee where
$|0\rangle$ is the vacuum state for the slave boson and the
pseudo-fermion operators, and for the singly-occupied state,
$|\sigma\rangle = \dbarsigma |e\rangle$. This assignment is very
different from $|e\rangle =\bbar |0\rangle$ and
$|\sigma\rangle=f^+_\sigma |0\rangle$ of the conventional
slave-boson theory. Note also that the expression for the empty
state is asymmetric in the interchange of $b_1$ and
$b_2$\cite{comment1}. Later we will see that this asymmetry is a
consequence of the finite bias across the dot.

In order to ensure that the new Hamiltonian expressed in terms of
slave bosons and pseudo-fermions has the same matrix elements as
the original $H$, we further require that $\langle \sigma |
\dbarsigma |e \rangle=1$ in the new representation. In meeting
this requirement, and (a)-(b) of Eq. (\ref{condition1}), we deduce
several relations: (1) $p=1+\kappa, q=1-\kappa$, (2)
$m^2=1-\kappa$, $n^2=\kappa$, (3) $\alpha^2 =\beta^2=1/2$. Now the
$d$-level operator in Eq. (\ref{dsigma}) is uniquely defined as
$d_\sigma = (\bbarl \flsigma +\bbarr \frsigma)/\sqrt{2}$. The
value of $\kappa$ will be determined later.

Hamiltonian in the slave-boson representation becomes \ba H_T &=&
{V_L \over \sqrt{2}} \sum_{k\in L,\sigma} \cbar_{Lk\sigma} (\bbarl
\flsigma +\bbarr \frsigma) + h.c. \nn &+& {V_R \over\sqrt{2}}
\sum_{k\in R,\sigma} \cbar_{Rk\sigma} (\bbarl \flsigma +\bbarr
\frsigma) + h.c.,\ea and \be H_d =-{\ed \over2} \sum_\sigma
(\fbarlsigma \flsigma + \fbarrsigma \frsigma ).\ee The complete
Hamiltonian together with the occupation constraints given in Eq.
(\ref{condition1})(a)-(b) becomes \ba H &=& H_L + H_R + H_d + H_T
\nn &+& \lambdal \left( \bbarl \bl + \sum_\sigma \fbarlsigma
\flsigma -1-\kappa \right) \nn &+&\lambdar \left( \bbarr \br +
\sum_\sigma \fbarrsigma \frsigma -1+\kappa
\right).\label{complete-H}\ea In the following we absorb
$\sqrt{2}$ in the definition of the tunnelling amplitudes, $V_L$
and $V_R$.

We have accomplished an {\it exact mapping} of the original
Hamiltonian, Eq. (\ref{original-H}), to an equivalent one, Eq.
(\ref{complete-H}), using the slave-boson theory valid in
$U\rightarrow\infty$ limit. Although the one-slave-boson theory of
AL also starts from an exact mapping of the original
Hamiltonian\cite{langreth}, the outcome of the mean-field analysis
may differ significantly depending on the type of mapping chosen.

One can carry out the stationary-state analysis of the ground
state of the above Hamiltonian in the manner of AL, solving for
the four parameters, $b_1$, $b_2$ and $\lambdal$ and $\lambdar$,
which requires \be b_{1(2)}^2 +\langle\sum_\sigma
f^+_{1(2)\sigma}f_{1(2)\sigma} \rangle = 1\pm\kappa,
\label{meanfield-1}\ee and \be 2\lambda_{1(2)} b_{1(2)}^2 +
\langle H_{T1(2)}\rangle = 0.\label{meanfield-2}\ee \widetext The
tunnelling part of the Hamiltonian factorizes, $H_T =
H_{T1}+H_{T2}$, with the $1$ and $2$ parts containing the $1$- or
$2$-dependent portion of the $d$-level operator. We replaced the
$b$-operators by their mean-field values $\bl$ and $\br$ that are
assumed to be real. Keldysh techniques are employed in evaluating
the averages\cite{meir-wingreen,langreth}. At zero temperature,
the mean-field conditions of Eqs. (\ref{meanfield-1}) and
(\ref{meanfield-2}) turn into

\ba {\pi\over2}(\bl^2 -\kappa) &=&{\gammal \over \gammal
+\gammar}\arctan \left ({\lambdal -\mu_L \over \bl^2
\Sigma_2}\right) +{\gammar \over \gammal+\gammar} \arctan \left
({\lambdal -\mu_R \over \bl^2 \Sigma_2}\right), \nn
{\pi\over2}(\br^2 +\kappa) &=&{\gammal \over \gammal
+\gammar}\arctan \left ({\lambdar -\mu_L \over \br^2
\Sigma_2}\right) +{\gammar \over \gammal+\gammar}\arctan \left
({\lambdar -\mu_R \over \br^2 \Sigma_2}\right),\label{mf-1}\ea and

\ba \ed = \gammal \ln \left(D^2 \over {(\lambdal-\mu_L)^2 + (\bl^2
\Sigma_2)^2} \right)+\gammar \ln\left(D^2 \over
{(\lambdal-\mu_R)^2 + (\bl^2 \Sigma_2)^2} \right), \nn \ed =
\gammal \ln \left(D^2 \over {(\lambdar-\mu_L)^2 + (\br^2
\Sigma_2)^2} \right)+\gammar \ln\left(D^2 \over
{(\lambdar-\mu_R)^2 + (\br^2 \Sigma_2)^2} \right). \label{mf-2}\ea
We have introduced $\gammal=V_L^2 \rho_L, \gammar =V_R^2 \rho_R$,
the width of the conduction band $D$, the self-energy $\Sigma_2 =
\pi (\gammal+\gammar)$, and re-defined $\lambda_{1(2)} -\ed +
b_{1(2)}^2 \Sigma_1$ as $\lambda_{1(2)}$ in the above expressions.
$\Sigma=\Sigma_1 +i\Sigma_2$ gives the complex self-energy of the
$d$-electron due to hybridization with the leads. We consider the
case $\ed\gg \lambda_{1(2)}, |\Sigma|$, as would be appropriate
for the Kondo regime.

\narrowtext Equations (\ref{mf-1})-(\ref{mf-2}) are the central
results of this paper. In the following we solve them for the case
of symmetric coupling, $\gammal=\gammar$, and in the limit of
infinite barrier to one of the leads, $\gammar \rightarrow 0$.

The bias voltage $\mu_R -\mu_L =V$ is taken to be positive (right
lead at higher chemical potential than the left lead). We look for
solutions in the Kondo limit $\bl^2 \approx 0 \approx \br^2$, and
$\lambdal$ and $\lambdar$ lying very close to either $\mu_L$ or
$\mu_R$. The case $\lambdal\approx\lambdar$ has a solution only
for a small bias, $V\ll b^2\Sigma_2$ and is not applicable at
large $V$. Hence we require that $\lambdal \approx \mu_L$, and
$\lambdar \approx \mu_R$. The other possibility, $\lambdal \approx
\mu_R$ and $\lambdar \approx \mu_L$, is obtained when the bias is
reversed, $V<0$.

The self-consistency equations are now reduced to (taking
$\gammal=\gammar$) \ba \bl^2 &\approx& b_0^2 /[1+\pi^2 (b_0^2
-\kappa')^2]^{1/2} \nn \br^2 &\approx& b_0^2 /[1+\pi^2 (b_0^2
+\kappa')^2]^{1/2},\label{bl-br}\ea where
$\kappa'=\kappa-(1/\pi)\arctan[V/b_0^2\Sigma_2 ]$, and

\be b_0^2 \Sigma_2 \approx {D^2\over max(V, b_0^2
\Sigma_2)}e^{-\pi \ed/\Sigma_2},\label{T-kondo-equation}\ee while
$\lambdal-\mu_L$ and $\lambdar-\mu_R$ satisfy

\be \lambda_{1(2)} -\mu_{L(R)} \approx \pi b_{1(2)}^2 \Sigma_2
\times \left[b_{1(2)}^2-\kappa'\right].\label{eq-24}\ee

Now we come to the determination of $\kappa$.  It is chosen to
minimize the energy $\langle H_T + H_d \rangle$ which, after
self-consistent relations Eqs. (\ref{meanfield-1}) and
(\ref{meanfield-2}) are taken into account, become $-{1\over2}\ed
(\bl^2 + \br^2)+const$. Other terms in the expectation value of
the Hamiltonian are insensitive to the choice of $\kappa$. Optimal
$\kappa$ is therefore one that maximizes $\bl^2 +\br^2$. Given the
expressions for $\bl^2$ and $\br^2$ obtained in Eq. (\ref{bl-br}),
it immediately follows that $\kappa'$ must be zero, or
$\kappa=\arctan[V/b_0^2\Sigma_2 ]$. Since the empty state
$|e\rangle$ is defined as $(\sqrt{1-\kappa}\bbarl\bbarr
+\sqrt{\kappa/2}(\bbarl)^2 ) |0\rangle$, we see that the
definition of the state itself depends on the applied bias! Put
another way our expression for the empty state contains a
variational parameter $\kappa$, which is chosen to maximize the
energy gain from the hybridization.

The mean-field solution demonstrates the existence of two Kondo
peaks since the $d$-electron spectral function is given by the sum
\be A_d (\omega)\approx {\bl^2 \over 2} A_1 (\omega) + {\br^2\over
2} A_2 (\omega)\label{e-spectral-function}\ee with the spectral
functions for the pseudo-fermions $f_1$ ($A_1 (\omega)$) and $f_2$
($A_2 (\omega)$) each producing a resonance at the chemical
potential of the lead\cite{cross-terms}.

The magnitude of $b_0^2\Sigma_2$ is the so-called Kondo
temperature, $T_K$. One can see from Eq. (\ref{T-kondo-equation})
that $T_K$ changes as the bias voltage is increased through the
threshold value, $V = D e^{-\pi \ed/2\Sigma_2}\equiv T_K^*$ in the
manner \ba T_K &\approx& T_K^* ~~~~~~~~~~ (V < T_K^*) \nn
 &\approx & (T_K^* )^2 /V ~~~ (V > T_K^*). \ea
Coleman {\it et al.}\cite{coleman} made a similar prediction for
the crossover behavior of the Kondo scale as the bias voltage is
increased through $\tstar$, based on the perturbative calculation
of the magnetic susceptibility. Our mean-field calculation
demonstrates the crossover behavior explicitly. The crossover
scale $\tstar$ is the Kondo temperature at equilibrium.

When tunnelling to one of the barriers is forbidden,
$\gammar\rightarrow 0$, we have as the solution $\kappa=0$, $b_0^2
\Sigma_2 = \tstar$, and $\lambdal =\lambdar \approx \mul$.
Therefore, the two spectral functions $A_1 (\omega)$ and
$A_2(\omega)$ merge to produce a single resonance peak at $\mul$,
as one would expect in the case of coupling to a single reservoir.
Further calculation (see next paragraph) shows that the split
resonance positions are close to the chemical potentials of each
reservoir as long as the tunnelling strengths, given by $\gammal$
and $\gammar$, are reasonably close to each other. As
$\gammar/\gammal$ becomes too small, eventually the resonance
level pinned to $\mur$ begins to float, and merge with $\mul$.

When $V\gg \tstar$, some understanding can be achieved for the
slightly asymmetric case, $\gammal+\gammar =\Gamma$,
$\gammal-\gammar =\Gamma \epsilon$ with $0<\epsilon\ll 1$. Here we
obtain that the Kondo temperatures scale differently for the two
resonances, with the $T_K$ for the stronger resonance near $\mu_L$
varying as $\tstar(\tstar/V)^{(1-\epsilon)/(1+\epsilon)}$, and for
the weaker resonance to the right reservoir, as
$\tstar(\tstar/V)^{(1+\epsilon)/(1-\epsilon)}$. In the highly
asymmetric case, {\it i.e.} $\gammal=\Gamma(1-\epsilon)$ and
$\gammar=\Gamma\epsilon$, we get the Kondo temperature scaling
$\tstar (\tstar/V)^{\epsilon/(1-\epsilon)}$ for both resonances.
Predicting such dependence of power-law exponents on the
tunnelling geometry would be very difficult based on other
calculation schemes, and illustrates the utility of SBMFT. It is
of course an interesting issue to see if experimentally such
power-law behavior indeed shows up.

In conclusion, we have introduced a new version of SBMFT for the
Kondo effect in a quantum dot that correctly reproduce the
equilibrium Kondo effect, and interpolates smoothly to the
non-equilibrium situation with a bias $V$ or to the case of
asymmetric coupling to the reservoirs. The introduction of two
sets of slave-bosons and pseudo-fermions may at first sight seem
to have doubled the number of orbitals in the dot, but on closer
inspection it is clear that the dimension of the Hilbert space for
the dot is strictly three, as ought to be the case for
$U\rightarrow\infty$. Rather, we believe this doubling of the
number of operators is tied to the presence of multiple reservoirs
with which the dot's electron can interact. Meanwhile, the
technical aspect of the present theory is completely
straightforward and may be easily generalized to understand Kondo
physics in other, more complex situations with multiple
dots\cite{langreth}, a phonon-coupled dot\cite{recent-experiment},
{\it etc.} Also of great interest is the role of the inelastic
processes, not considered in the present mean-field theory, in
reducing the Kondo scale $T_K$\cite{non-equil-kondo}. Future work
will address these issues.

We acknowledge helpful discussions with Kicheon Kang, Tae-Suk Kim,
Hyun-Woo Lee, and Heung-Sun Sim. The completion of the paper took
place during participation at the APCTP Focus Program ``Quantum
Effects in Nanosystems". This research is supported by Korea
Research Foundation Grant (KRF-2002-003-C00042) and by Korean
Science and Engineerging Foundation (KOSEF) through Center for
Strongly Correlated Materials Research (CSCMR) at Seoul National
University.

\widetext


\begin{references}
\bibitem{N-leads} In the coupling of $N$ leads to the dot, one can
presumably write $d_\sigma$ as a sum, $\sum_{i=1}^N \alpha_i b^+_i
f_{i\sigma}$ ($\sum_i \alpha_i^2 =1$). We restrict ourselves to
$N=2$ in this paper.
\bibitem{comment1} One can equally well have written $|e\rangle = (m \bbarl \bbarr +
{n\over\sqrt{2}}(\bbarr)^2)|0\rangle$ without a loss of
generality.
\bibitem{cross-terms} One can show that the cross terms such as
$\langle \fbarlsigma \frsigma \rangle$ are $O(b_0^2)$ smaller than
the direct terms shown in Eq. (\ref{e-spectral-function}).
\end{references}
\end{document}